\documentclass[journal=jacsat,manuscript=article]{achemso}

\usepackage{chemformula} 
\usepackage[T1]{fontenc} 
\usepackage{lipsum}
\usepackage{graphicx}
\graphicspath{{./images_main/}}
\usepackage[caption=false]{subfig}
\usepackage{amsmath}
\usepackage{verbatim}
\usepackage{float}
\usepackage{tabularx}
\usepackage{threeparttable}
\usepackage{dcolumn}
\usepackage{bm}
\usepackage{color,epsfig,multirow}
\usepackage{array}
\usepackage{hyperref}
\usepackage{algpseudocode}
\usepackage{algorithmicx,enumerate,algorithm}
\usepackage{booktabs}
\usepackage{threeparttable}
\usepackage{dcolumn}
\usepackage{geometry}

\author{Teng Zhao}
\affiliation{School of Mathematical Sciences, Shanghai Jiao Tong University,\\ Shanghai, 200240, China}

\author{Shuangliang Zhao}
\affiliation{State Key laboratory of Chemical Engineering and School of Chemical Engineering, \\
	East China University of Science and Technology, Shanghai, 200237, China \\
	Guangxi Key Laboratory of Petrochemical Resource Processing and Process Intensification Technology and School of Chemistry and Chemical Engineering, Guangxi University, Nanning, 530004, China
}

\author{Shenggao Zhou}
\email{sgzhou@sjtu.edu.cn}
\affiliation{School of Mathematical Sciences, MOE-LSC, CMA-Shanghai and Shanghai Center for Applied Mathematics, Shanghai Jiao Tong University, Shanghai, 200240, China}

\author{Zhenli Xu}
\email{xuzl@sjtu.edu.cn}
\affiliation{School of Mathematical Sciences, MOE-LSC, CMA-Shanghai and Shanghai Center for Applied Mathematics, Shanghai Jiao Tong University, Shanghai, 200240, China}

\title[An \textsf{achemso} demo]
  {How Thermal Effect Regulates Cyclic Voltammetry of Supercapacitors}
\abbreviations{IR,NMR,UV}

\begin{document}

\begin{abstract} 
Cyclic voltammetry (CV) is a powerful technique for characterizing the electrochemical properties of electrochemical devices. During charging-discharging cycles, the thermal effect can have a profound impact on its performance. However, existing theoretical models cannot clarify such intrinsic mechanism and often give poor prediction. Herein, we propose an interfacial model for the electro-thermal coupling, based on fundamentals in non-equilibrium statistical mechanics. By incorporating molecular interactions, our model shows a quantitative agreement with experimental measurements. The integral capacitance shows a first enhanced then decayed trend against the applied heat bath temperature. Such a relation is attributed to the competition between electrical attraction and Born repulsion via dielectric inhomogeneity, which was not well understood in previous models. In addition, as evidenced in recent experimental CV tests, our model predicts the non-monotonic dependence of the capacitance on the bulk electrolyte density. This work demonstrates a potential pathway towards next-generation thermal regulation of electrochemical devices.

\end{abstract}

\section{TOC Graphic}
\begin{figure}
	\centering
	\includegraphics[height = 5.0cm, width = 7.5cm]{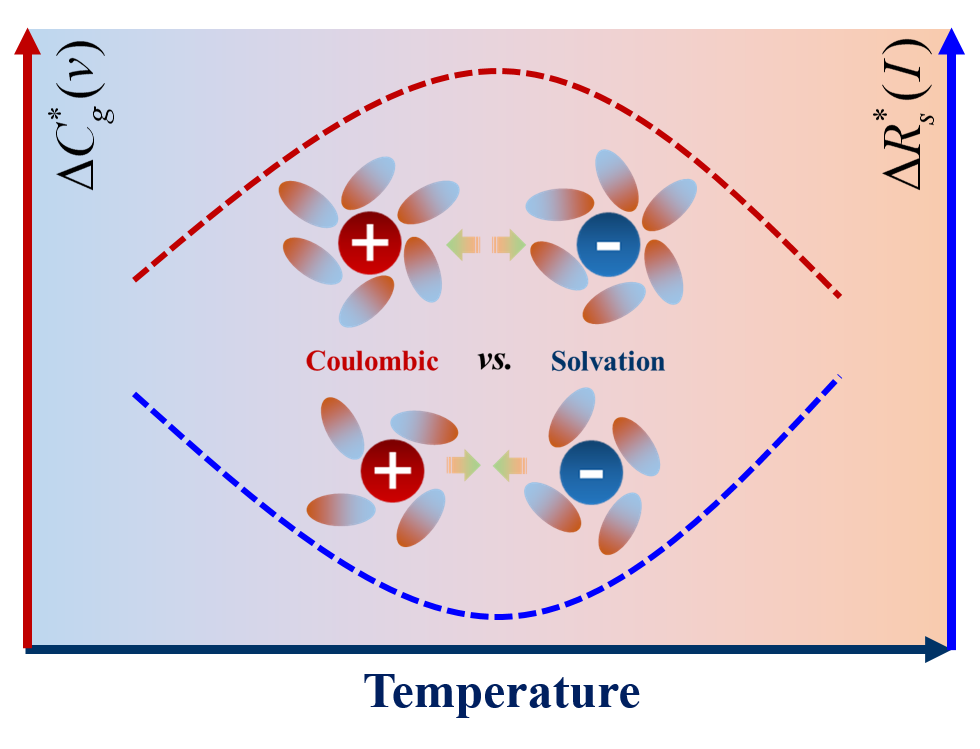}
\end{figure}

Recently, supercapacitors \cite{supercapacitor_Science, Supercapacitors_ACS_EL_1, Supercapacitors_ACS_EL_2} have drawn significant attention in electrical energy storage applications, in particular those requiring rapid charging-discharging operations. Owing to higher power density \cite{high_Capa_Angew}, larger cycle efficiency \cite{cycle_effe_Carbon}, longer lifetime \cite{supercapacitor_long_cycle_life} and other advantageous features, supercapacitor-based micro/nano-devices have not only taken as an ideal device for energy storage, but also greatly been adopted for many related fields both in fundamental research and engineering applications \cite{review_supercapacitor_2}, such as energy harvesting \cite{heat_harvesting_Science, blue_energy_harvesting_Nature_review, Supercapacitors_ACS_EL_3}, purification and ion separation \cite{EDLC_app}.   \par 

Most of the experimental measurements of commercial supercapacitors is based on 
a powerful technique, termed cyclic voltammetry (CV) \cite{CV_book}, in which the current density is determined in response to an externally applied periodic potential. Specifically, with a given scan rate, the applied voltage gradually increases in a charging phase, and then decreases in a  discharging phase with a reversed scan rate. Electric energy is stored via the the adsorption of ions into electric double layers, and it is released accompanied by the ion dissolution back to the bulk. Due to the simplicity and convenience in manipulation, CV tests are widely utilized to determine the electrical properties of supercapacitors.  \par 

Heat generation \cite{heat_EDLC_1, Pilon_JPS_2014} during the operation has become a major concern, since these electrochemical devices are usually cycled under high current density. The thermal effect has great influence on many aspects related to supercapacitor performance, such as cell aging \cite{cell-aging}, dendrite growth \cite{dendrite-growth} and internal resistance increase \cite{thermal_effect_capacitor}. However, the thermal effect can also be used to regulate and enhance the energy storage performance \cite{SI_1, SI_2, SI_3}, providing an attractive way to realize the directional regulation and effective management of such electrochemical devices. In addition, the supercapacitor performance can also be influenced by many other factors such as pH \cite{pH_Carbon}, electrolyte size and valence \cite{eps_wetting_AICHE}, asymmetric electrode material components \cite{asymmetric_electrode}, light excitation effect \cite{light_effect}, magnetic effect \cite{magnetic_effect}, and the thermal effect can also have coupling effect on these factors to regulate the power and energy densities, especially for extreme environmental temperatures \cite{low_temp_supercap}. For example, under low temperature, the charge and mass transfer of electrolytes are generally suppressed \cite{low_temp_AEM}, resulting in performance decay. Qin \emph{et al.} \cite{magnetic_effect} synthesized a multi-responsive healable supercapacitor with magnetic hydrogel-based electrodes, and they found that the hydrogel exhibits fast healing properties triggered by the magneto-thermal induced interfacial reconstructions. Therefore, comprehensive understanding of the thermal effect on electric performance during charging-discharging processes is significant and desirable, in order to improve the safety, performance and long-term stability of supercapacitors. \par 

The thermal effect can be studied via both experimental and theoretical approaches. Most experimental investigations measure the macroscopic averaged thermal behavior of supercapacitors, typically assuming that the heat generation rate remains uniform across the whole device \cite{Cg_4}. Alternatively, various theoretical methods are developed for the charge dynamics and thermal transport inside supercapacitors, such as the microscopic ``stack-electrode'' model and its equivalent circuit \cite{current_density_Lian_PRL}, and a thermodynamically-consistent model via an energetic variational approach \cite{SZhou_JPS_2022}. In general, theoretical models are based on macroscopic description of mass and energy conservation, neglecting detailed molecular interactions. Such treatment is questionable for describing ionic species confined in the micro/nano-scale porous medium, in which the interfacial effect becomes remarkable \cite{Heat_transfer_Xi_CES, JPS_Zhao_2023}.   \par 

To understand the thermal effect on CV performance inside supercapacitors from a molecular perspective, herein we propose a microscopic heat transfer model coupled with interfacial ionic structures. Based on non-equilibrium statistical mechanics, the theoretical framework demonstrates that it can unravel the underlying mechanism of local temperature variation effect on CV tests. Different from the macroscopic heat transfer models, the proposed  model is rigorously derived from the Liouville equation, and molecular interactions are inherently encoded in such first-principle derivation. We further show that the non-uniform ionic structure has a significant impact on the local temperature evolution. Several approaches regulating the EDL structure and thermal transport are systematically investigated as well. In addition, our model is able to predict experimental data on integral gravimetric capacitance, especially the non-monotonic transition where the state-of-the-art theoretical approach fails.  \par

\begin{figure}[htbp]
	\centering
	\includegraphics[scale = 0.58]{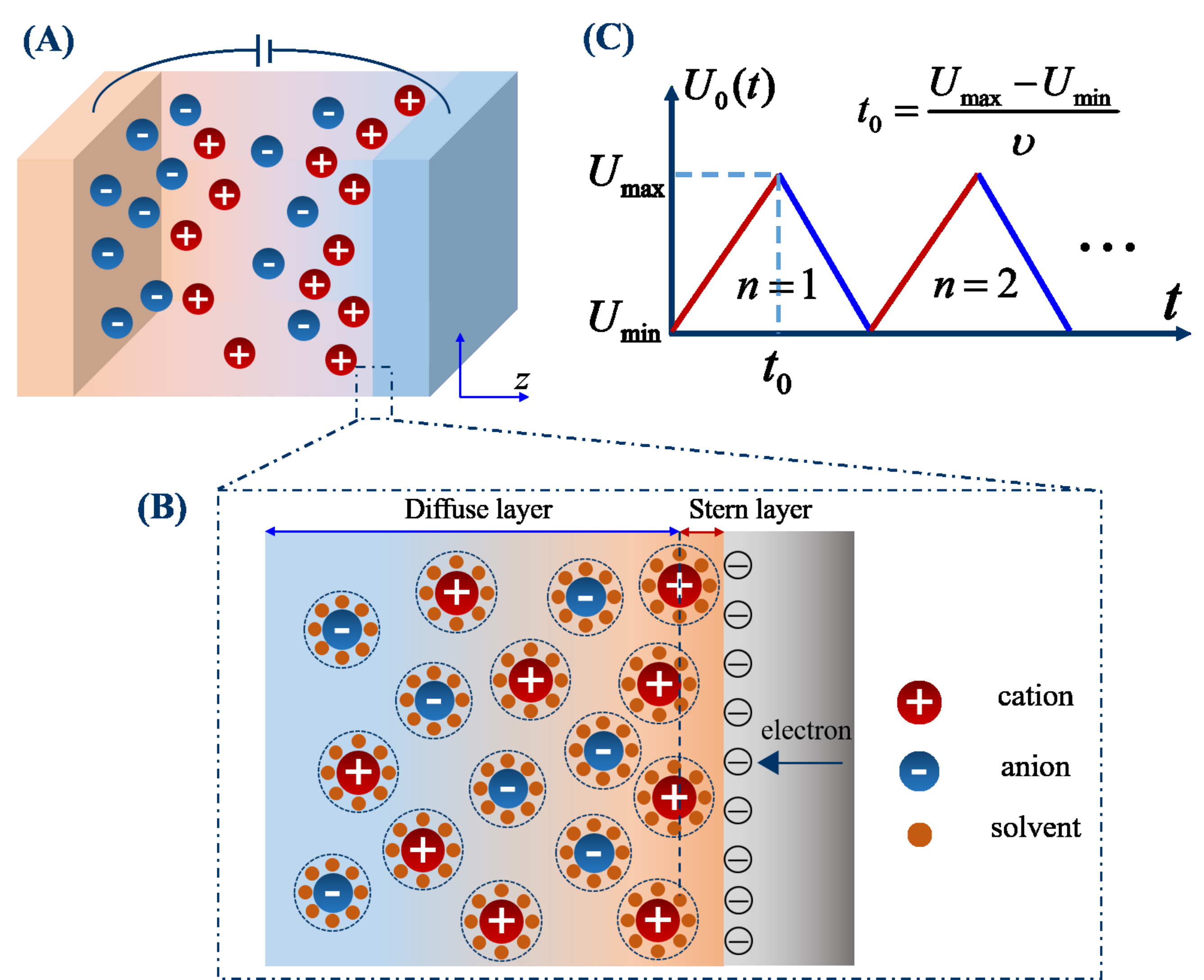}
	\caption{Schematic diagram of cyclic voltammetry processes in a supercapacitor with separation $ H $: (A) charging-discharging processes connected with an external circuit to provide gating voltages on the electrodes; (B) Electric double layer (EDL) structure of hydrated cations and anions near a planar electrode; (C) Temporal evolution of the applied gating voltage with a given scan rate $ v $. During the charging/discharging cycles, ions are adsorbed to/released from the EDL structure near the electrode surface, accompanying the reversible heat generated/converted mainly at the interfacial zone.}
	\label{fig_1}
\end{figure}

A schematic diagram describing the cyclic voltammetry process in a supercapacitor is shown in Fig.~\ref{fig_1}, where the capacitor is modeled as two parallel hard walls with separation $ H $. The electrolyte confined inside the capacitor is described by the restricted primitive model (RPM). In RPM, both cations and anions are represented by charged hard spheres, while the solvent is implicitly modeled as a continuum with a relative dielectric coefficient \cite{Lian_ACS_Nano}. It is noted that the dielectric coefficient can be regulated by the local fluid density \cite{eps_rho_depend_Qing}, surface wettability (hydrophobicity of the electrode) \cite{eps_wetting_AICHE, eps_wetting_SZhao}, external field effect \cite{magnetic_effect} (magnetic, electric) and many other factors. These effects are previously well studied, while the local thermal effect is rarely investigated in depth. In this work, the impact of local temperature variation on solvent dielectric coefficient is taken into account via a both spatially and temporally dependent function $ \epsilon_{r} = \epsilon_{r} \left[ T(z;t) \right] $. In addition, it is assumed that both cations and anions have the same diameter $ \sigma $ with opposite valences $ Z_+ = -Z_- = 1$ without loss of generality. Due to geometrical symmetry of the system, dynamical properties, such as the local density and temperature profiles, only vary in the direction perpendicular to the wall surface, viz. the $ z $-direction.   \par 

A whole CV test of supercapacitors can be separated into periodic charging-discharging cycles. First, in the charging process, the capacitor is connected to an external circuit shown in Fig.~\ref{fig_1} (A), and then the voltages applied on the positively-charged ($ U_{+} $) and negatively-charged ($ U_{-} $) electrodes drive ion adsorption to form EDL structures shown in Fig.~\ref{fig_1} (B). In the charging phase (see red lines in Fig.~\ref{fig_1} (C)), the voltage gradually promotes with a positive scan rate $ v $. On the contrary, in the discharging process, the applied voltage decays with a negative scan rate (blue lines in Fig.~\ref{fig_1} (C)), then the adsorbed ions start to dissolve back into the surrounding bulk environment, resulting in the collapse of the EDL structure. During a charging-discharging cycle, the corresponding electric current is recorded and shown as the IV curve for further electrochemical analysis.   \par

The evolving gating voltage with a given scan rate $ v $ is of a saw-tooth wave shape between a minimum value $ U_{min} $ and a maximum value $ U_{max} $, and is defined as:~
$$
	U_{0}(t) = \begin{cases}
		U_{min} + v t  & 2(n-1)t_{0} \leq t \leq (2n-1)t_{0}, \\
		U_{max} - v[t - (2n-1)t_{0}]  & (2n-1)t_{0} \leq t \leq 2nt_{0},
	\end{cases} 
$$
\noindent where $ t_{0} = (U_{max} - U_{min}) / v $ represents a half period of a charging-discharging cycle, and $ n = 1, 2, \cdots $ is the cycle number. The gating voltage and surface charge density evolutions during a cycle can be seen in Fig.~S1 in the supporting information (SI).   \par 

\begin{figure}[htbp]
	\centering
	\includegraphics[scale = 0.32]{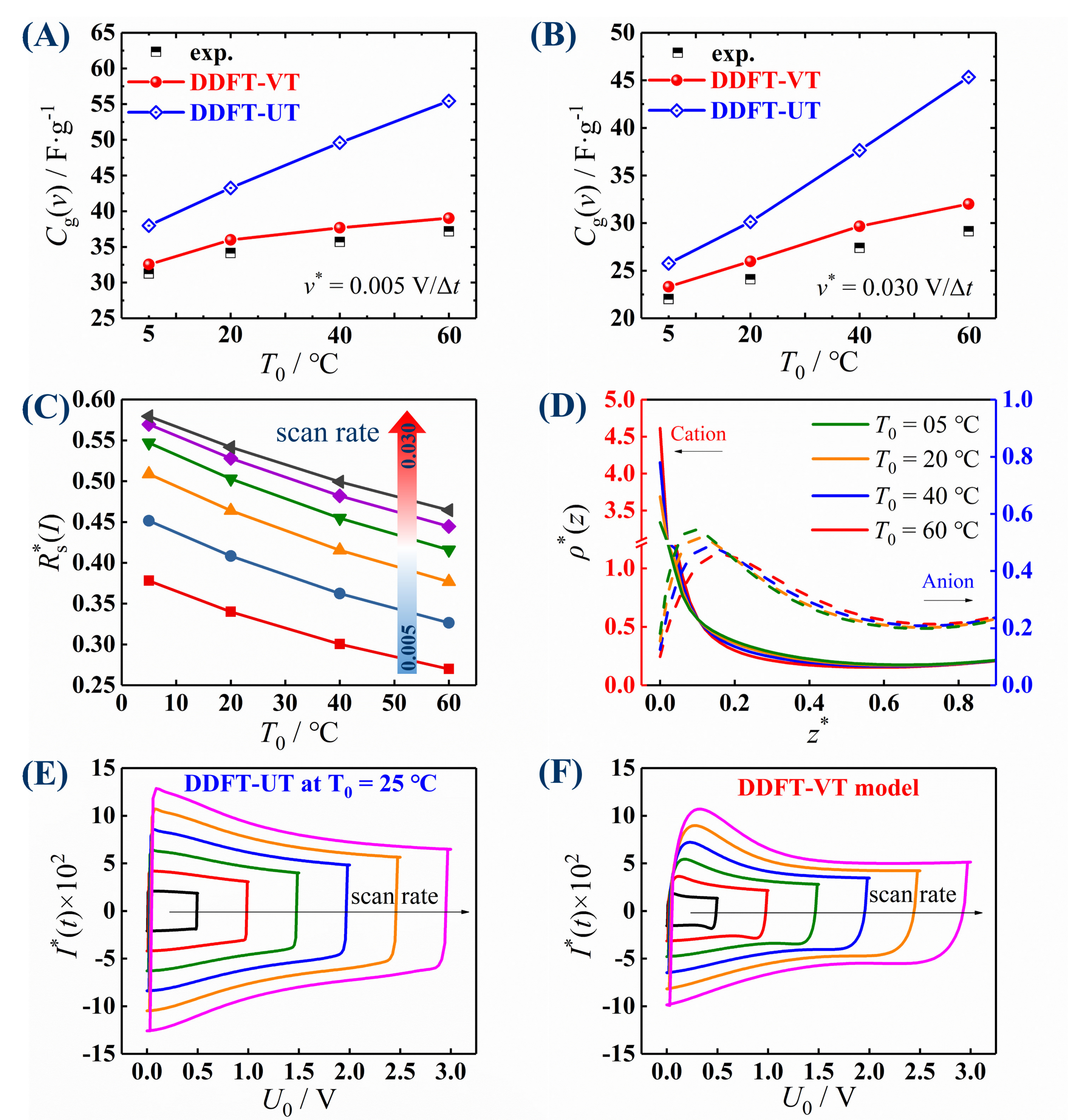}
	\caption{Comparison of experimental measurement (square scatter) and theoretical prediction for the integral gravimetric capacitance with various scan rates: (A) $ v^{*} = 0.005 $; (B) $ v^{*} = 0.005 $; the red and blue lines represent CV tests calculated by a DDFT model with a uniform temperature $T_0= 25 $ \textcelsius (labeled as \textit{DDFT-UT}), and the proposed model in this work with varying temperature profile (labeled as \textit{DDFT-VT}), respectively. (C) The dependence of internal resistance of the capacitor on temperature with various scan rates. (D) Density profiles of ions under various system temperatures; (E) and (F) represent cyclic voltammogram with respect to scan rates without/with local temperature variations, respectively. The minimum and maximum gating voltages applied on the electrode surface are $ U_{\rm min} = \pm 0.0 $ V and $ U_{\rm max} = \pm 0.5 $ V, respectively. The dimensionless bulk ionic density is $ \rho_b^{*} = 0.3 $, and the system width is $ H = 4 \sigma $, where the ion diameter $ \sigma = 0.5 $ nm. }
	\label{fig_2_new}
\end{figure}

Now we employ the proposed dynamical density functional theory (DDFT) with varying temperature (DDFT-VT) model to investigate the impact of thermal effects on the cyclic voltammetry of a supercapacitor. CV processes with/without local temperature evolution are performed for validation of the proposed model. To quantitatively assess the electric performance during the charging-discharging cycles, we calculate the integral gravimetric capacitance and the internal resistance. It is observed from Fig.~\ref{fig_2_new} (A, B) that, for both low and high scan rates, the theoretical calculation with the local temperature variation has good quantitative agreement with the experimental data, showing much better performance than that with a uniform temperature model (labeled as DDFT-UT at 25 \textcelsius). It is also found that the capacitance promotes with the increase of temperature. From a molecular perspective, such a conclusion is attributed to the fact that molecular interactions are enhanced in high temperature. As a result, ions near the electrode form a denser EDL structure with the temperature promotion, as shown in Fig.~\ref{fig_2_new} (D). We also give CV curves under different scan rates from $ v^{*} = $ 0.005 to 0.030 shown in Fig.~S2, and for each scan rate, the CV curve is enlarged with the increase of temperature, indicating that the energy storage performance is enhanced.   \par  

In addition, the internal resistance is also taken as an important quantity to characterize the supercapacitor performance, and it can be calculated from the IR drop during the charging-discharging transition. For example, Likitchatchawankun \emph{et al.} \cite{Cg_4} experimentally investigated supercapacitors consisted of two identical activated carbon electrodes with 1 M $ \rm Pyr_{14}TFSI $. It is found that with the increase of ambient temperature, the measured internal resistance decreases. As demonstrated in Fig.~\ref{fig_2_new} (C), with the increase of the temperature, the calculated internal resistance also decreases significantly, being consistent with the experimental measurement. Such a phenomenon indicates that the electro-diffusion inside the capacitor enhances \cite{CV_exp_Pilon}, as temperature increases. As a result, the EDL formation becomes stronger and results in a higher charge storage capacity.   \par 

We also compare the CV curves without/with local temperature variations in Fig.~\ref{fig_2_new} (E, F). For each given condition, when the applied scan rate is rather small (e.g., $ v^{*} = 0.005 $), the enclosed CV curve is like a quasi-rectangular shape, indicating that the sign of electric current is immediately reversed upon the reversal of the external potential sweep. Such a rectangular-like CV curve usually characterizes an ideal double-layered capacitance behavior \cite{CV_curve_shape_Carbon}. On the other hand, when the scan rate is gradually enhanced (e.g., $ v^{*} = 0.030 $), a deviation from such a rectangular shape can be observed from the CV curves. Due to the much faster voltage elevation under larger scan rates, the ions are strongly adsorbed to the electrode surface driven by electrical attraction. As a result, the EDL structure becomes much denser than that under small scan rates. Under such circumstances, the charge stored in the capacitor is strongly enhanced. Different from CV curves with fixed temperature (Fig.~\ref{fig_2_new} (E)), Fig.~\ref{fig_2_new} (F) shows that the current gradually promotes during the charging phase and the CV curve becomes smoother than that in (F). In addition, it is noted that there exits a ``hump'' in CV curve \cite{CV_exp_Pilon}. Such a phenomenon is attributed to the local variation of dielectric profile induced by the local temperature change, the ion adsorption has become saturated near the electrode surface before reaching the maximum potential. We also plot CV curves under four different heat bath temperatures (Fig.~S3), and the hump can all be observed. In addition, the CV test under cold environment is also investigated, which is shown in Fig.~S4. Due to the suppression of ion adsorption, the integral gravimetric capacitance is greatly decayed.    \par   

\begin{figure}[!h]
	\centering
	\includegraphics[scale = 0.22]{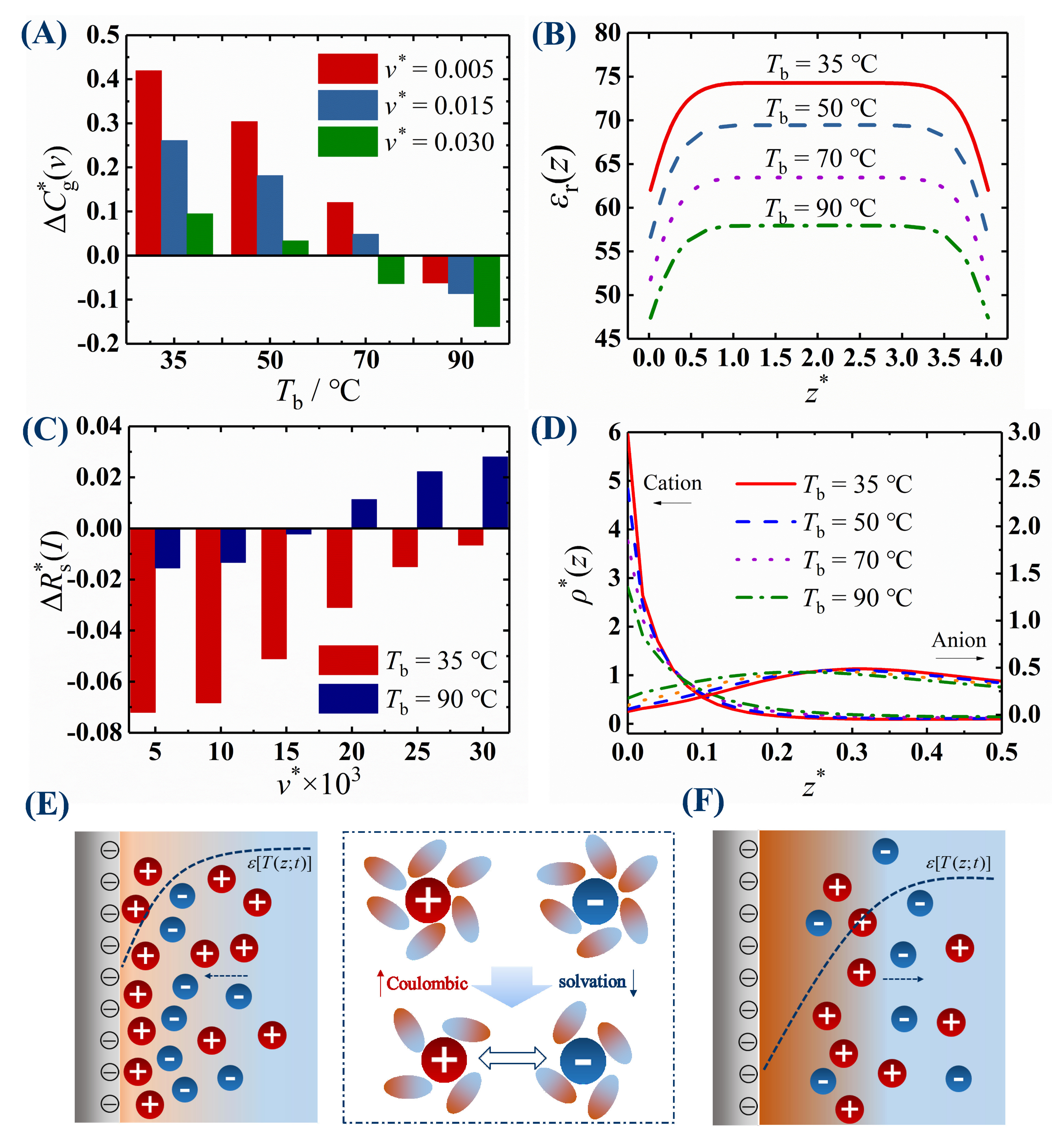}
	\caption{(A) Integral gravimetric capacitance difference with respect to the heat bath temperature $ T_{\rm b} $; (B) dielectric coefficient profiles; (C) internal resistance difference under different scan rates, with the red and blue bars representing $ T_{\rm b} = 35 $ and $ T_{\rm b} = 90 $ \textcelsius, respectively; (D) density profiles of cations and anions near the electrode surface; (E) and (F) represent the illustrations of molecular mechanism of thermal effect on EDL structure under low and high temperatures.}
	\label{fig_3_new}
\end{figure}

To further clarify the thermal effect, the difference between integral gravimetric capacitance with and without applied heat bath (also known as ambient temperature), viz. $ \Delta C_{g}^{*}(v) = C_{g}(T_{\rm b}) - C_{g}(T_{0})$,
and the corresponding internal resistance difference $ \Delta R_{s}^{*}(I) $ are shown in Fig.~\ref{fig_3_new}. For a given scan rate in Fig.~\ref{fig_3_new} (A) (e.g., $ v^{*} = 0.005 $, red bar), it is found that for low ambient temperature (e.g., $ T_{\rm b} = 35 $ \textcelsius), the capacitance difference is greater than zero, indicating that the applied heat bath enhances charge storage capability. When the temperature is promoted, from $ T_{\rm b} = $ 35 to 70 \textcelsius, such enhancement due to dielectric decay near the electrode surface is weakened. If the temperature continues to increase (e.g., $ T_{\rm b} = 90 $ \textcelsius), the capacitance difference is then reversed, which means $ \Delta C_{g}^{*}(v) $ is now less than zero. From Fig.~\ref{fig_3_new} (B), it is observed that the decrement of dielectric coefficient near the electrode surface becomes much stronger with the increase of heat bath temperature. From a molecular perspective, the thermal effect on dielectric variation can be attributed to that the vibration of solvent molecules promotes with the increase of temperature, and it impedes rotational dipole moment. As a result, the dielectric coefficient decreases \cite{dielectric_decay_CPL, dielectric_decay_Science}. From the internal resistance difference shown in Fig.~\ref{fig_3_new} (C), it is found that the charge transfer resistance drops more under $ T_{\rm b} = 35 $ compared to $ T_{\rm b} = 90 $, especially for small scan rate. Consequently, the contact density of counterions decreases near the electrode surface as the temperature promotes, while the coions slightly increases (Fig.~\ref{fig_3_new} (D)). Therefore, the net charge accumulation gradually decays with increasing heat bath temperature. To better understand the molecular mechanism, we plot schematic diagrams shown in Fig.~\ref{fig_3_new} (E-F) for more detailed clarification. First, due to the dielectric decay near the electrode surface caused by the applied heat bath, the electrostatic screening effect is weakened while the electrical attraction is enhanced. As a result, more ions are adsorbed near the electrode surface and make the EDL structure become denser. However, if the dielectric decrement continues to be enhanced (at higher temperature), the ionic species are excluded from the electrode surface, and the EDL structure can no longer be maintained by the electrostatic interaction and then it becomes rather loose. Such first-enhanced-then-weakened phenomenon indicates that the capacitance can be regulated by the local dielectric environment variation near the electrode surface. The non-monotonic transition of capacitive energy storage performance with respect to the environment temperature has also become the focus in relevant researches \cite{Nature_letters, annual_review_high_temp, Science_PoreSize_2006}, especially for Coulombic efficiency failure in high-temperature applications \cite{Nature_HuangXY}.   \par 

\begin{figure}[h!]
	\centering
	\includegraphics[scale = 0.35]{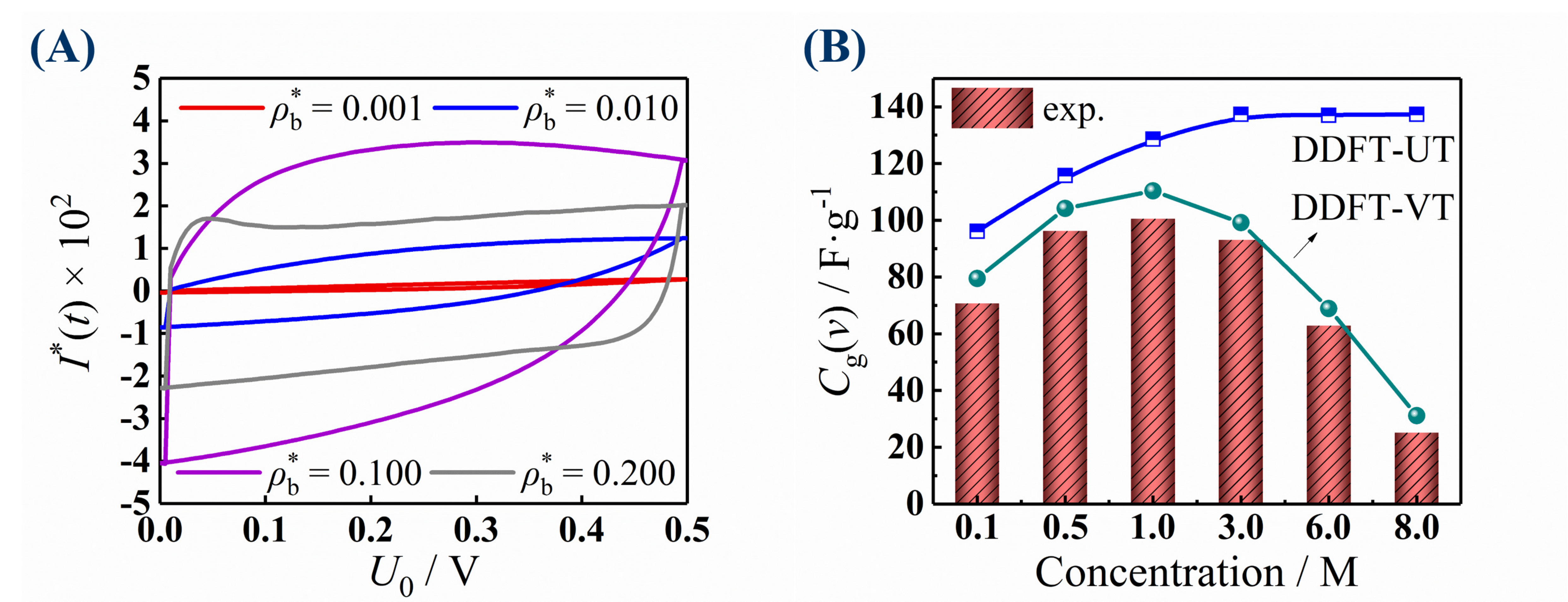}
	\caption{(A) Cyclic voltammogram under different bulk densities; (B) comparison of integral gravimetric capacitance between theoretical calculations and experimental measurements with the same system parameters as in the work~\cite{CV_bulk_density}. }
	\label{fig_9}
\end{figure}

The bulk density of ions has great influence on EDL structure and thermal transport inside capacitors, as shown in Fig.~\ref{fig_9} (A). When the bulk density is rather small (e.g., $ \rho_{b}^{*} = 0.001 $), the CV curve is compressed like a needle, which means there are too few ions to form an EDL structure. Therefore, the electric current density is rather small. With the increase of bulk density from $ \rho_{b}^{*} = 0.005 $ to $ \rho_{b}^{*} = 0.100 $, the CV curve gradually enlarges due to the enhancement of EDL adsorption by electrostatic interactions. However, if the density continues to increase ($ \rho_{b}^{*} = 0.200 $), the electric current density decays instead, leading to a shrinkage of the CV curve. As a result, the integral gravimetric capacitance shown in Fig.~\ref{fig_9} (B) first promotes then decays with respect to the bulk density. Such a non-monotonic relation is due to the competition between electrical interactions and Born repulsion due to the dielectric decrement effect. To be specific, when the bulk density is rather small, the electrostatic interactions dominate and are enhanced with the increase of bulk density, and thus the charge accumulation is promoted. If the bulk density continues to increase, the dielectric coefficient is then suppressed near the electrode surface, leading to repulsion of ions by dielectric inhomogeneity. As a result, the integral capacitance first boosts and then decays as the bulk density increases. It should also be noted that such a non-monotonic transition predicted by our model has also been validated in recent experimental CV tests by Chen \emph{et al.}~\cite{CV_bulk_density} (Fig.~\ref{fig_9} (B)). In contrast, the DDFT-UT model with uniform fixed temperature cannot predict such non-monotone transition, indicating that the thermal effect is crucial to theoretical prediction of experimental capacitance of supercapacitors.   \par 

It should also be noted that the interfacial reactions occurred on the electrode surfaces have a great impact on thermal transfer inside electrochemical devices. In general, the Butler-Volmer equation \cite{BV_equation_JPS}, accounting for the relation between current density and surface over-potential, is often used for describing phenomenological electrode kinetics. Recently, Tang \emph{et al.} \cite{DRxDFT_CES} has developed a dynamic reaction density functional theory (DRxDFT) based on the molecular collision theory to deal with reaction-diffusion coupling at nanoscale. In our future work, we will extend the DRxDFT framework so that the electrode reaction-heat transfer coupling can be achieved.   \par

In conclusion, a first-principle DDFT-VT model based on non-equilibrium statistical mechanics is developed to investigate the thermal effect on cyclic voltammetry of supercapacitors. The microscopic equations for the description of local fluid structure and local temperature with molecular interactions are derived based on the fundamentals of non-equilibrium statistical mechanics. With an experimental formula between permittivity and local temperature, we apply the proposed electro-thermal coupling model to study the thermal effect on CV tests. The theoretical model is validated by compared with experimental data, presenting a significant improvement over the original DDFT with a uniform temperature. It is found that with the increase of heat bath temperature, the capacitance difference is first enhanced then suppressed, which is consistent with recent experimental studies. It is demonstrated that such a transition is attributed to the competition between electrical attractions and Born repulsion due to permittivity inhomogeneity. Finally, we study the effect of bulk electrolyte density on the EDL structure and dielectric profiles. It is found the capacitance shows a non-monotonic dependence on an increasing bulk density, agreeing with experimental data quantitatively. This work offers a molecular perspective into the thermal regulation on supercapacitor performance during CV tests. Our model is expected to be promising in the rational design and effective thermal management of electrochemical devices.   \par

\textbf{Methods.} The microscopic dynamical governing equations describing the evolution of local temperature and ion structure inside capacitors are derived from the Bogoliubov--Born--Green--Kirkwood--Yvon (BBGKY) hierarchical equation, which describes the temporal evolution of the probability distribution function based on the Liouville equation \cite{Kreuzer_NESM}. It is noted that all dynamics concerned in this work, including local ion structure and local temperature profiles, is encoded in the distribution function \cite{SF_SZhao, ZHAO2021116623}. The detailed derivations are presented in the SI. Herein, the theoretical framework and characterization methods are summarized as follows.  \par 

First, we introduce the governing equations for mass and heat transfer in CV tests. By integrating both sides of the first equation of the BBGKY hierarchy with respect to the moment $ \mathbf{p} $, one has the following dynamical governing equation for the evolution of local ionic structure $ \rho(\mathbf{r}_{i}; t) $ for the \textit{i}-th species \cite{SF_SZhao, ZHAO2021116623, AIChE_Zhao}: 
\begin{equation}
	\begin{aligned}
		\frac{\partial}{\partial t} \rho(\mathbf{r}_{i}; t) = \nabla_{\mathbf{r}_{i}} & \cdot \left\{ \frac{D_{i}}{k_B T(\mathbf{r}_{i}; t)} \rho(\mathbf{r}_{i}; t) \left[ \nabla_{\mathbf{r}_{i}} \frac{\delta F\left[ \rho(\mathbf{r}_{i}; t) \right]}{\delta \rho(\mathbf{r}_{i}; t)} \right] \right\} \\ 
		\quad &+ \nabla_{\mathbf{r}_{i}} \cdot \left\{ \frac{D_{i}}{k_B T(\mathbf{r}_{i}; t)} \nabla_{\mathbf{r}_{i}} \cdot \left[ \rho(\mathbf{r}_{i}; t) \mathbf{u}(\mathbf{r}_{i}; t) \otimes \mathbf{u}(\mathbf{r}_{i}; t) \right] \right\}, 
	\end{aligned}   \label{4} 
\end{equation}

\noindent where $ \mathbf{u}(\mathbf{r}_{i}; t) $ represents the local fluid velocity, $ T(\mathbf{r}_{i}; t) $ is the local temperature profile, $ k_B $ is the Boltzmann constant, and the self-diffusion coefficient $ D_{i} $ is related to the friction coefficient by $ (\gamma_{i})^{-1} = D_{i} / k_B T $. Here $ F \left[ \rho(\mathbf{r}_{i}; t) \right] $ represents the total Helmholtz free-energy functional, for which the variational derivative with respect to the local density gives the local chemical potential $ \mu(\mathbf{r}_{i}; t) $. $ F \left[ \rho(\mathbf{r}_{i}; t) \right] $ can be decomposed into the intrinsic Helmholtz free-energy functional $ F_{\rm int} \left[ \rho(\mathbf{r}_{i}; t) \right] $ and the external free-energy functional $ F^{\rm ext} \left[ \rho(\mathbf{r}_{i}; t) \right]  $. The former consists of the ideal-gas part $ F^{\rm id} \left[ \rho(\mathbf{r}_{i}; t) \right]$ and an excess part $ F^{\rm ex} \left[ \rho(\mathbf{r}_{i}; t) \right] $, describing non-interaction contributions and molecular interactions among particles \cite{DFT_review_SZhao, DFT_AIChE_Wu}, respectively.   \par 

Multiplying both sides of the first BBGKY hierarchy equation by $ \left[ \mathbf{p} - \mathbf{p}(\mathbf{r}_{i}; t) \right]^2 / 2 m_{i} $ and then integrating the resulting equation with respect to $ \mathbf{p} $, one obtains
\begin{equation}
	\begin{aligned}
		\frac{\partial T(\mathbf{r}_{i}; t)}{\partial t} =& \frac{2}{3k_B \rho(\mathbf{r}_{i}; t)} \nabla_{\mathbf{r}_{i}} \cdot \left\{  \lambda_{i} \nabla_{\mathbf{r}_{i}} T(\mathbf{r}_{i}; t) \right\}  \\
		&- \frac{T(\mathbf{r}_{i}; t)}{\rho(\mathbf{r}_{i}; t)} \nabla_{\mathbf{r}_{i}} \cdot \left\{ \frac{D_{i}}{k_B T(\mathbf{r}_{i}; t)} \rho(\mathbf{r}_{i}; t) \left[ \nabla_{\mathbf{r}_{i}} \frac{\delta F\left[ \rho(\mathbf{r}_{i}; t) \right]}{\delta \rho(\mathbf{r}_{i}; t)} \right] \right\} + Q_s, 
	\end{aligned}  \label{5} 
\end{equation}
\noindent with $ Q_s $ being the heat source and $ \lambda_{i} $ representing the thermal conductivity. The source term for heat generation is introduced to describe the Joule heating effect: $ Q_s = K_{e}|\mathbf{E}|^2/\rho_b c_p $, where $ K_{e} $ denotes the electric conductivity of ionic liquids, $ \mathbf{E} $ is the electric field, $ \rho_b $ represents the bulk density, and $ c_p $ is the heat capacity \cite{JPS_Zhao_2023}.  \par 

For the confined charged hard-sphere system under consideration, the Helmholtz free-energy functional generally contains  contributions from the external potential part, the ideal-gas part and the excess part \cite{DFT_review_SZhao}. The total excess Helmholtz free-energy functional is expressed as
$$ F^{ex}(\mathbf{r}; t) = F^{ex}_{hs}(\mathbf{r}; t) + F^{ex}_{C}(\mathbf{r}; t) + F^{ex}_{el}(\mathbf{r}; t) + F^{ex}_{Born}(\mathbf{r}; t), $$ \noindent where $ F^{ex}_{hs}(\mathbf{r}; t) $, $ F^{ex}_{C}(\mathbf{r}; t) $, $ F^{ex}_{el}(\mathbf{r}; t) $ and $ F^{ex}_{Born}(\mathbf{r}; t) $ represent the contribution of hard-sphere repulsion \cite{MFMT_Yu}, Coulomb interactions, electrostatic correlation \cite{MSA_Blum} and Born hydration energy of ions \cite{Born_Xu_SIAM_2018, Born_Xu_SIAM_2021}, respectively. Taking the first-order functional variation of the Helmholtz free energy with respect to the density, we obtain the local electrochemical potential $$ \mu_{i}(\mathbf{r}; t) = \mu_{i}^{\rm id}(\mathbf{r}; t) +\mu_{i}^{\rm hs}(\mathbf{r}; t) + \mu_{i}^{\rm C}(\mathbf{r}; t) + \mu_{i}^{\rm el}(\mathbf{r}; t) + \mu_{i}^{\rm Born}(\mathbf{r}; t) + V_{0}^{\rm ext}(\mathbf{r}), $$
\noindent where $ V_{0}^{\rm ext}(\mathbf{r}) $ is non-electrical external potential to ions originating from the electrode-ion interaction. The mean electrical potential $ \psi(\mathbf{r}; t) $ solves the Poisson equation 
\begin{equation}
	\nabla \cdot \left\{ \epsilon_{r}\left[T(\mathbf{r}; t), \rho_{s} \right] \nabla \psi(\mathbf{r}; t) \right\} = - \frac{e_{0}}{\epsilon_{0}} \sum_{i = +, -} Z_{i} \rho(\mathbf{r}_{i}; t), \label{6}  
\end{equation}
\noindent where $ \epsilon_{0} $ represents the permittivity of free space, $ e_{0} $ is the elementary charge, and $ \epsilon_{r} $ represents the local dielectric profile that depends on temperature and bulk density.
The following semi-empirical relation fitting the experimental data is utilized \cite{eps_temp_depend_PRL_Roij}:
\begin{equation}
	\begin{aligned}
		\epsilon_{r} \left[T(\mathbf{r}; t); \rho_{s}\right]  = 87.88 - 0.39T(\mathbf{r}; t) &+ 0.00072T^{2}(\mathbf{r}; t) \\ 
		\times (1.0 - 0.2551\rho_{s} &+ 0.05151\rho_{s}^{2} - 0.006889\rho_{s}^{3}),
	\end{aligned}  \label{7} 
\end{equation}
\noindent where $ T(\mathbf{r}; t) $ is in Celsius degree and $ \rho_{s} $ is the bulk density in molar. It should be emphasized that the dielectric coefficient of liquid water at atmospheric pressure decreases monotonically with temperature increase, resulting in a promotion of the Bjerrum length.  \par 

The characterization methods of CV performance are summarized as follows. The supercapacitor cell integral gravimetric capacitance $ C_{g} $ (in F/g) can be evaluated by integrating the area enclosed by the CV curve that plots measured current $ I(t) $ versus  imposed external potential $ U_{0} $ on the electrode under a given scan rate $ v $ \cite{Cp_int_1}, viz.,
\begin{equation}
	C_g(v) = \frac{1}{M(U_{max} - U_{min})} \oint \frac{I}{2 v} \mathrm{d}U,  \label{8}  
\end{equation}

\noindent where $ U_{max} $ and $ U_{min} $ represent the maximum and minimum potential during a scanning process, respectively. Their difference $ U_{max} - U_{min} $ is known as the potential window. Also, $ M $ is the mass of active material loaded into both electrodes. The electric current density $ I(t) $ in the external circuit measured in CV experiments is defined as \cite{current_density_Lian_PRL}: $ I(t) = dQ(t) / dt $, where $ Q(t) $ represents the time-dependent surface charge density and can be calculated by \cite{Q_t_Jiang}
\begin{equation}
	Q(t) = Q_{eff}(t) - \frac{\beta e_{0}^2 \partial \psi(z; t)}{4 \pi l_{B} \partial z} \bigg|_{z = H/2}.  \label{9}
\end{equation}
Here $ \psi(z; t) $ is the local electrical potential solved by the Poisson equation, $ l_{B} $ is the Bjerrum length, and $ \beta = 1 / k_{B}T_r $, where $ T_r $ is the absolute temperature of the system. The $ Q_{eff}(t) $ represents the effective surface charge density (a.k.a. net charge accumulation), which is defined as:
$$ Q_{eff}(t) = -\int_{0^{+}}^{H/2} \rho_{e}(z;t) \mathrm{d}z, $$ where $ \rho_{e}(z;t) = \sum_{i = +, -} e_{0} Z_{i} \rho_{i}(z;t) $ is the net charge density.  \par 

In addition, the internal resistance (or DC resistance) $ R_{s} $  was determined from the voltage drop at charging-discharging transitions according to \cite{Cg_4}
\begin{equation}
	R_{s}(I) = \frac{U_{s}(t_{c}^{+}) - U_{s}(t_{c}^{-})}{2 I},  \label{10} 
\end{equation}

\noindent where $ U_{s}(t_{c}^{+}) $ and $ U_{s}(t_{c}^{-}) $ are the potentials across the capacitor cell at the end of the charging phase and immediately after the beginning of the discharging step, respectively.   \par

\begin{acknowledgement}

This work is supported by the National Natural Science Foundation of China (Nos. 12171319, 12071288 and 22178072). S. Zhou acknowledges the support from Shanghai Science and Technology Commission 21JC1403700. Z. Xu acknowledges the support from the Science and Technology Commission of Shanghai Municipality (grant Nos. 20JC1414100 and 21JC1403700). T. Zhao acknowledges the support from China Postdoctoral Science Foundation (No. 2022M712055). The authors also acknowledge the support from the HPC center of Shanghai Jiao Tong University.

\end{acknowledgement}

\begin{suppinfo}

The following files are available free of charge. \par 
\begin{itemize}
	\item S.1 Detailed derivation of the theoretical framework
	\item S.2 Evaluation of Helmholtz free-energy functional
	\item S.3 Supplementary graphs
\end{itemize}

\end{suppinfo}

\bibliography{CV_ThermalEffect_ACS_MainText_Ref}

\end{document}